\documentclass[10pt,amssymb,superscriptaddress,showkeys,twocolumn,pra]{revtex4-1}
\usepackage{amsmath,latexsym}
\usepackage[english]{babel}
\usepackage{graphicx}
\usepackage{amsmath}
\usepackage[pdftex, pdftitle={Article}, pdfauthor={Author}]{hyperref} 
\usepackage{natbib}
\usepackage{graphicx}
\usepackage{dcolumn}
\usepackage{bm}
\usepackage{color}

\begin{document}

\setlength{\abovedisplayskip}{3pt}
\setlength{\belowdisplayskip}{3pt}
\setlength{\abovedisplayshortskip}{3pt}
\setlength{\belowdisplayshortskip}{3pt}

\preprint{APS/123-QED}

\title{Spontaneous Symmetry Breaking In Nonlinear Binary Periodic Systems }

\author{Ruihan Peng}
\thanks{These authors contributed equally to this work}
\affiliation{State Key Laboratory of Advanced Optical Communication Systems and Networks, School of Physics and Astronomy, Shanghai Jiao Tong University, Shanghai, China.}
\affiliation{School of Physics and Astronomy, Shanghai Jiao Tong University, Shanghai 200240, China.}

\author{Qidong Fu}
\thanks{These authors contributed equally to this work}
\affiliation{State Key Laboratory of Advanced Optical Communication Systems and Networks, School of Physics and Astronomy, Shanghai Jiao Tong University, Shanghai, China.}
\affiliation{School of Physics and Astronomy, Shanghai Jiao Tong University, Shanghai 200240, China.}

\author{Yejia Chen}
\affiliation{School of Physics and Astronomy, Shanghai Jiao Tong University, Shanghai 200240, China.}

\author{Weidong Luo}
\affiliation{School of Physics and Astronomy, Shanghai Jiao Tong University, Shanghai 200240, China.}

\author{Changming Huang}
\email{hcm123\_2004@126.com}
\affiliation{Department of Physics, Changzhi University, Changzhi, Shanxi 046011, China}

\author{Fangwei Ye}
\email{fangweiye@sjtu.edu.cn}
\affiliation{State Key Laboratory of Advanced Optical Communication Systems and Networks, School of Physics and Astronomy, Shanghai Jiao Tong University, Shanghai, China.}
\affiliation{School of Physics and Astronomy, Shanghai Jiao Tong University, Shanghai 200240, China.}


\begin{abstract}
{Spontaneous symmetry breaking (SSB) occurs when modes of asymmetric profile appear in a symmetric, double-well potential, due to the nonlinearity of the potential exceeding a critical value. In this study, we examine SSB in a periodic potential where the unit cell itself is a symmetric double-well, in both one-dimensional and two-dimensional periodic systems. Using the tight-binding model, we derive the analytical form that predicts the critical power at which SSB occurs for both 1D and 2D systems. The results show that the critical power depends significantly on the quasi-momentum of the Bloch mode, and as the modulus of momentum increases, the SSB threshold decreases rapidly, potentially dropping to zero. These analytical findings are supported by numerical nonlinear eigenmode analysis and direct propagation simulations of Bloch modes.

}
\end{abstract}

\maketitle


\section{\label{sec:level1}introduction}

Spontaneous symmetry breaking (SSB) is a key concept in various branches of physics~\cite{malomed2013spontaneous,10.21468/SciPostPhysLectNotes.11}. It refers to a situation that, due to the effect of nonlinearity, a symmetric ground state reshapes itself into an asymmetric one, expressed as an imbalance in the population of the modes despite the fact that the underlying potential is symmetric. SSB is observed in various nonlinear systems, such as those involving liquid crystals~\cite{vanakaras1998tilt}, quantum dots~\cite{yannouleas1999spontaneous}, Bose-Einstein condensates~\cite{jackson2004geometric,matuszewski2007spontaneous,wang2009two,waclaw2009tuning}, and nonlinear optics~\cite{malomed2013spontaneous}.

In the context of Bose-Einstein condensation, the SSB phenomenon may appear in a natural setting, such as the double well potential. The experimental realization of the SSB~\cite{albiez2005direct,gati2006realization} and the SSB effects for matter wave solitons~\cite{gubeskys2007symmetric,matuszewski2007spontaneous,trippenbach2008spontaneous,salasnich2010competition,zegadlo2016spontaneous} in such double-well potentials have been reported. 
In optics, the optical structure provides a fertile platform to explore novel symmetry-breaking states, such as systems with single resonators~\cite{kaplan1982directionally,del2017symmetry,copie2019interplay}, coupled resonators~\cite{ghosh2024controlled,pal2024linear,mai2024spontaneous}, double-well sites~\cite{brazhnyi2011spontaneous,li2012nonlinear,chen2013switch,kevrekidis2005spontaneous}, 
coupled nonlinear waveguides~\cite{mak1997solitons,mak1998asymmetric,kevrekidis2005spontaneous,driben2006all,dror2011symmetric,li2012two}, 
dual-core fibers~\cite{wright1989solitary,akhmediev1993novel,albuch2007transitions,mayteevarunyoo2011spontaneous},
dual-cavity lasers~\cite{heil2001chaos,hamel2015spontaneous,malomed2015symmetry}, nano- and micro-scaled structures~\cite{christ2008symmetry,Huang:11,yang2014feedback,zhang2019symmetry},
and various types of waveguide arrays~\cite{bulgakov2011symmetry,wang2011spontaneous,Dolinina:20}.
SSB has also been studied in non-conservative~\cite{miroshnichenko2011nonlinearly,achilleos2012dark,xu2021spontaneous} and fractional dimensional systems~\cite{li2019pt,li2020symmetry,li2021symmetry}.

The above mentioned works demonstrate that the fundamental symmetric structures support localized, symmetry ground states, and adding nonlinearity to these structures disrupts the symmetry of the ground state, provided that the strength of the nonlinearity surpass a critical threshold. 
In that case, symmetric and asymmetric nonlinear states with identical eigenvalues (such as chemical potential in BECs or propagation constant in optics) coexist, albeit with different number of atoms or power. 

On the whole, SSB is a fundamental effect caused by the interplay between nonlinearity and symmetric structures. Therefore, the symmetry breaking behaviors can be strongly influenced by the structural geometry and parameters. The majority of the  aforementioned works is centered on investigating the properties of SSB in the simplest double configurations~\cite{brazhnyi2011spontaneous,li2012nonlinear,chen2013switch,mak1997solitons,mak1998asymmetric,kevrekidis2005spontaneous,driben2006all,dror2011symmetric,li2012two,wright1989solitary,akhmediev1993novel,albuch2007transitions,mayteevarunyoo2011spontaneous,heil2001chaos,hamel2015spontaneous,malomed2015symmetry,christ2008symmetry,Huang:11,yang2014feedback,zhang2019symmetry}. Additionally, SSB has been explored in three-core linearly coupled triangular configurations, such as optical fibers~\cite{akhmediev1994soliton} and Bragg lattices~\cite{gubeskys2004solitons}. Note that, however, all these studies focus on  one-dimensional, finite systems, with double- or triple-well potentials, the SSB in periodic systems with binary potential in unit cells, as well as that in the two-dimensional systems haven not been studied
yet.  

In this work, with reference to the one-dimensional Su-Schrieffer-Heeger (SSH) model and the two-dimensional honeycomb lattice, we study the symmetry breaking of Bloch modes in nonlinear binary periodic systems.  
We find the SSB point, at which asymmetric nonlinear Bloch modes bifurcates from symmetric Bloch modes, depends crucially on the quasi-momentum $k$ of the considering Bloch modes. Specifically, in the one-dimensional SSH model, the critical power of SSB decreases as the quasi-momentum $k$ varies from $0$ to $\pi/L$, with $L$ being the lattice constant. Similarly, in the two-dimensional honeycomb lattice,  we observe a similar trend as $k_x$ and $k_y$ vary along the direction from $\Gamma$ to $M$ and then to $K$. In both binary periodic lattice systems, quasi-momentum plays an essential role in the SSB of nonlinear Bloch modes, which is a degree of freedom missing in the previously studied,  finite optical structures.



%


\section{\label{sec:level2}One-dimensional SSH model}
\subsection{\label{sec:level2a} Model and Analytical Analysis}
\begin{figure}[htbp]
	\centering
	\includegraphics[width=1.0\columnwidth]{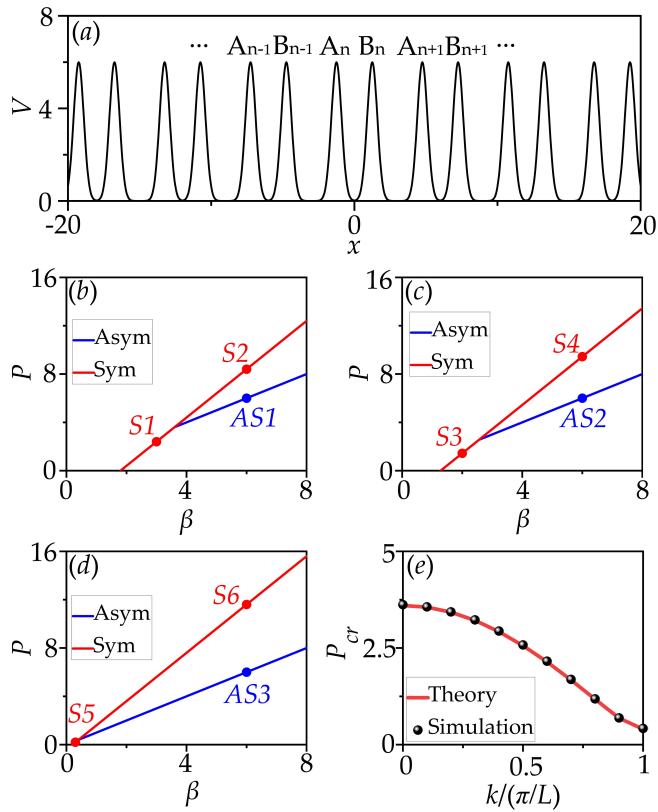}
	\caption{Panel (a) presents the schematic of the waveguide array aligned in the transverse direction. There are two waveguides, $A$ and $B$, in a unit cell. Panels (b), (c) and (d) show the power per unit cell of the symmetric (red line) and asymmetric Bloch mode (blue line) at $k=0$ , $k=\pi/2L$ and $k=\pi/L$, respectively. Red and blue dots correspond to the mode profiles presented in Fig. 2. (e) The critical power $P_\text{cr}$ curve of different Bloch momentum $k$. Here, $c_1=1, c_2=0.8$.}
	\label{fig1}
\end{figure}

We consider the propagation of a light beam along the $z$ axis of a focusing Kerr medium, using a one-dimensional SSH model, as illustrated in Fig.~\ref{fig1}(a). For the sake of convenience, we designate waveguide $A$ and waveguide 
$B$ as the two waveguides in the unit cell. In this binary periodic waveguide array, the coupling constants exhibit 
alternation, which we denote as $c_1$ and $c_2$ respectively.
By employing the tight-binding approximation and considering only nearest-neighbor waveguide coupling, one finds the propagation of light in the waveguide array  can be described by the following set of nonlinear coupled equations:
\begin{equation}\label{eq1}
\begin{aligned}
&i\frac{\partial{Q_{a,n}}}{\partial{z}}=-c_{1}Q_{b,n}-c_{2}Q_{b,n-1}-|Q_{a,n}|^{2} Q_{a,n}, \\
&i\frac{\partial{Q_{b,n}}}{\partial{z}}=-c_{1}Q_{a,n}-c_{2}Q_{b,n+1}-|Q_{b,n}|^2 Q_{b,n}.
\end{aligned}
\end{equation}
Here, $Q_{a,n}$ and $Q_{b,n}$ represent the optical field distributions of waveguide $A$ and waveguide $B$ in the $n$th unit cell, respectively. 
In this description, $c_1$ and $c_2$ characterizes the intracell and intercell coupling coefficients, respectively. We aim to find and study spatially periodic solution to Eqs.~(\ref{eq1}), thus, according to the Bloch theorem, the field amplitudes in the $n$-th unit cell, $Q_{a,n}$ and $Q_{b,n}$, are related to those in neighboured  $Q_{a,n+1}$ and $Q_{b,n-1}$ through $Q_{a,n}=Q_{a,n+1}e^{-ikL}$ and $Q_{b,n}=Q_{b,n-1}e^{ikL}$, respectively, where $k \in (-\pi/L,\pi/L]$ represents the Bloch momentum and $L$ is the lattice constant. Substituting these relations into Eqs.~(\ref{eq1}), we obtain the following equations:
\begin{equation}\label{eq2}
\begin{aligned}
&i\frac{\partial{Q_{a,n}}}{\partial{z}}=(-c_{1}-c_{2}e^{ikL})Q_{b,n}-|Q_{a,n}|^2 Q_{a,n}, \\
&i\frac{\partial{Q_{b,n}}}{\partial{z}}=(-c_{1}-c_{2}e^{-ikL})Q_{a,n}-|Q_{b,n}|^2 Q_{b,n}.
\end{aligned}
\end{equation}


 The stationary nonlinear solutions of Eqs.~(\ref{eq2}) can be expressed in the form of $Q_{a,n}=\varphi_{a}e^{i \beta z}$ and $Q_{b,n}=\varphi_{b}e^{i \beta z}$, where the wave functions $\varphi_{a}$ and $\varphi_{b}$, as well as the propagation constant $\beta$, satisfy  the following set of nonlinear equations:
\begin{equation}\label{eq3}
\begin{aligned}
&\beta \varphi_{a}=(c_{1}+c_{2}e^{ikL})\varphi_{b}+|\varphi_{a}|^2 \varphi_{a}, \\
&\beta \varphi_{b}=(c_{1}+c_{2}e^{-ikL})\varphi_{a}+|\varphi_{b}|^2 \varphi_{b}.
\end{aligned}
\end{equation}

Looking at Eqs.~(\ref{eq3}), one immediately finds that it recovers back to a simple double-well potential when the intercell coupling $c_2$ is set to zero. Thus, in binary periodic systems, due to the existence of quasi-momentum $k$, interactions between different unit cells introduces an additional Bloch phase factor $e^{ikL}$. This leads to an \textit{effective} intracell coupling coefficient $C_{\text{eff}}=c_1+c_2e^{ikL}$,  being the sum of the \textit{authentic} intracell coupling coefficient $c_1$ and the additional inter-cell term $c_2 e^{ikL}$. This effective coupling coefficient $C_{\text{eff}}$ directly reflects inter-cell interaction in periodic systems, and as it strongly depends on the value of quasi-momentum $k$, one expects that it will significantly alter the threshold power for SSB in periodic systems. In particular, for Bloch waves with $kL=\pi$, the effective coupling coefficient is reduced to its minimum value $C_{\text{eff}}=c_1-c_2$, leading to the lowest threshold power for symmetric modes to occur, as  will be shown below.

We aim to solve Eqs.~(\ref{eq3}) analytically. As mentioned, this equation recovers back to a simple double-well potential by setting the intercell coupling $c_2$ to be zero, and in that case, Eqs.~(\ref{eq3}) show that both $\varphi_a$ and $\varphi_b$ can be chosen as real numbers. For periodic systems with nonzero $c_2$, however, the equations require that $\phi_a$ and $\phi_b$ cannot both be real, and a phase difference between them must exist. Thus, to be specific, we assume $\varphi_{a}=\phi_{a}$ and $\varphi_{b}=\phi_{b}e^{i \theta}$, where $\phi_{a}$ and $\phi_{b}$ are real numbers, and $\theta$ is their phase difference. Substituting these wavefunctions into Eqs.~(\ref{eq3}) yields the following equations: 
\begin{equation}\label{eq4}
\begin{aligned}
&\beta \phi_a  =  (c_1+c_2e^{ikL})e^{i\theta}\phi_b + \phi_a^3, \\
&\beta \phi_b  =  (c_1+c_2e^{-ikL})e^{-i\theta}\phi_a + \phi_b^3.
\end{aligned}
\end{equation}

Taking into account both $\phi_a$ and $\phi_b$ are real numbers, the balance of Eqs.~(\ref{eq4}) requires $(c_1+c_2e^{ikL})e^{i\theta}$ and $(c_1+c_2e^{-ikL})e^{-i\theta}$ both real too. Due to their complex conjugate relation with each other, namely, $(c_1+c_2e^{ikL})e^{i\theta}=(c_1+c_2e^{-ikL})e^{-i\theta}$, one immediately finds that the phase shift between two waveguides satisfies $\tan \theta=-c_2\sin kL/(c_1+c_2\cos kL)$. Therefore, it becomes evident that the phase difference $\theta$ depends crucially on the Bloch momentum $k$.

In the following, for brevity, we set $(c_1+c_2e^{ikL})e^{i\theta}=(c_1+c_2e^{-ikL})e^{-i\theta}=C$, or $C=\sqrt{c_1^2+c_2^2+2c_1c_2\cos kL}$,
and from Eqs.~(\ref{eq4}) we have:
\begin{equation}\label{eq5}
\begin{aligned}
&(\beta+C)(\phi_{a}-\phi_{b})=\phi^{3}_{a}-\phi^{3}_{b}, \\
&(\beta-C)(\phi_{a}+\phi_{b})=\phi^{3}_{a}+\phi^{3}_{b}.
\end{aligned}
\end{equation}
Now it is ready to distinguish the symmetric, antisymmetric, and asymmetric modes by considering different combinations of $\phi_a, \phi_b$ and $\theta$. 
Since we are concerned with the symmetry breaking from the fundamental mode, namely, from the symmetric modes, we will not consider the antisymmetric modes in this work, although they can be solved in a similar way as the way shown below.  By definition, symmetric modes have $\phi_{b}=\phi_{a}$ (while antisymmetric modes have $\phi_{b}=-\phi_{a}$), and asymmetric modes, in contrast, feature $\phi_{b}\neq \pm\phi_{a}$, i.e., unequal light intensity in the two sites. Aimed at this asymmetric solution, Eqs.~(\ref{eq5}) would be tranformed as:
\begin{equation}\label{eq6}
\begin{aligned}
&\beta+C=\phi_{a}^{2}+\phi_{b}^{2}+\phi_{a}\phi_{b}, \\
&\beta-C=\phi_{a}^{2}+\phi_{b}^{2}-\phi_{a}\phi_{b}.
\end{aligned}
\end{equation}
Now we define the power in the waveguide $A$ and $B$ as $P_a=|\phi_{a}|^2$ and $P_b=|\phi_{b}|^2$, respectively, 
and with this Eqs.~(\ref{eq6}) take the following form:
\begin{equation}\label{eq7}
\begin{aligned}
&\beta+C=P_{a}+P_{b}+\sqrt{P_{a}P_{b}}, \\
&\beta-C=P_{a}+P_{b}-\sqrt{P_{a}P_{b}}.
\end{aligned}
\end{equation}
Eqs.~(\ref{eq7}) yield the power distribution of nonlinear asymmetric modes ($\phi_b \neq \phi_a$)  as, 
\begin{equation}\label{eq8}
\begin{aligned}
&P_{a,b}=\phi^{2}_{a,b}=\frac{\beta \pm \sqrt{\beta^2-4C^2}}{2}.
\end{aligned}
\end{equation}
Thus, interestingly, we find that in a periodic binary system, the symmetry breaking occurs when the nonlinear eigenvalue $\beta$ exceeds $2C$,  or equivalently,  when threshold power  of the symmetric mode exceeds $P_\text{cr}$,
\begin{equation}\label{eq9}
P_{cr}\equiv P_{a}+P_{b}=\beta= 2C=2\sqrt{c_1^2+c_2^2+2c_1c_2\cos kL}.
\end{equation}
Eqs.~(\ref{eq9}) clearly indicates that, in the binary periodic nonlinear systems with fundamental symmetric modes being periodic Bloch functions, the SSB point in the symmetric modes depends crucially on the Bloch momentum $k$. This is in sharp contrast to the isolated binary systems, where the SSB point is exclusively defined by the parameters of the underlying systems. Thus, while Bloch mode at different $k$ all exhibits symmetry breaking after power increases to some critical value, this critical powers drops significantly from $k=0$ for which $P_{cr}=3.6$ [Fig. 1(b)], to $k=\pi/2L$ for which $P_{cr}=2.56$ [Fig. 1(c)], and to $k=\pi$ for which $P_{cr}=0.4$ [Fig. 1(d)]. Considering $k$ in the half of the first Brillouin zone $k \in [0,\pi/L]$, $P_{cr}$ decreases monotonically with the Bloch momentum $k$, and achieves the lowest SSB breaking threshold for Bloch modes at the border of the Brillouin zone where $k=\pi/L$ [Fig. 1(e)]. 


\begin{figure}[htbp]
	\centering
	\includegraphics[width=1.0\columnwidth]{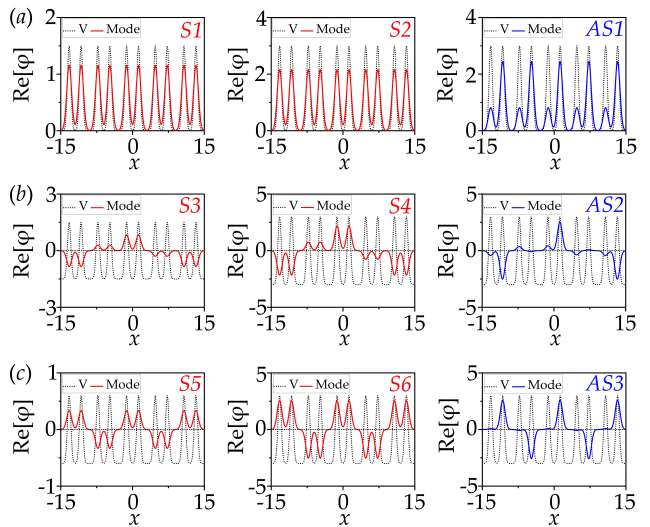}
	\caption{Examples of profiles of symmetric (left two columns) and asymmetric (right column) mode corresponding to the solid points   marked in Figs.~\ref{fig1}(b),~\ref{fig1}(c) and \ref{fig1}(d).
	$k=0$ in panels (a), $k=\pi/2L$ in panels (b), and $k=\pi/L$ in panels (c). The waveguide array is shown as the black lines in each panel.}
	\label{fig2}
\end{figure}

\subsection{\label{sec:level2b}Numerical Eigen-analysis}

While we have derived an analytical expression for the symmetry-breaking power threshold $P_{cr}$, Eqs.~(\ref{eq9}),
the nonlinear eigenmodes of Eq.~(\ref{eq3}) can be numerically computed using a standard iteration method starting from a guess symmetric solution. Once the solution for a specific $\beta$ is obtained, it is used as the initial guess for determining the solution corresponding to the subsequent, new $\beta$. To ensure the convergence of the iteration algorithm, a sufficiently small step in  $\beta$ was taken. 

The result for $k=0$ is presented in Fig.~1(b) and Fig. 2(a). It is observed that as $\beta$ increases, the power of the solution increases but remains symmetric, as shown in Fig.~2(a, left), until it arrives at a critical power $P_\text{cr}=3.6$. Beyond this point, a new family of solution branches out from the symmetric family, whose profiles are asymmetric [Fig.~2(a, right)]. However, it should be noted that the symmetric family continuous beyond the branching point, see Fig.~2(a, middle) for the example of a high-power symmetric solution. Note that,  for illustrative purposes, here and below, the field amplitudes in all pertinent figures are multiplied  by the waveguide mode-field profiles.

A Similar situation occurs for nonlinear solutions at different quasi-momentum $k$, however, with different critical power at which the asymmetric modes branch off. For comparison, Fig.~1(d) shows the case for $k=\pi/L$. In this case, the phase of the nonlinear solutions flips between adjacent unit cell, for both symmetric and asymmetric modes. 
However, the asymmetric modes branches from the symmetric family at a significantly lower power of $P_{cr}=0.4$. The influence  of the quasi-momentum $k$ on the $P_{cr}$ is thus apparent. Figure 1(e) presents the dependence of $P_{cr}$ on $k$, which shows that the numerical results perfectly agree with the analytical predictions, i.e., Eqs.~(\ref{eq9}).
\label{eq10}

 \subsection{\label{sec:level2c} Propagation Simulation of SSB in 1D Binary Systems}

The above-mentioned analytical and numerical eigen-analysis is further confirmed through direct light propagation simulation. To fully capture the possible physical effects such as the long-range coupling beyond the nearest-next-neighboured waveguides and the radiations loss, we here use the continuous equation in the following form,

\begin{equation} \label{eq10}
i \frac{\partial \Psi}{\partial z}=-\frac{1}{2} \frac{\partial^2 \Psi}{\partial x^2}-V(x)\Psi-|\Psi|^2\Psi
\end{equation}

Here, $x$ is the (continuous) transverse coordinate, $z$ is the propagation coordinate. The term  $V(x)=p\sum_{i} [e^{-\frac{(x-x_{2i})^2}{a^2}}+e^{-\frac{(x-x_{2i+1})^2}{a^2}}]$ describes the periodic waveguide arrays consisting of Gaussian waveguides with radius $a$ and amplitude $p$. In the framework of the SSH-like model we are considering, each unit cell consists of two Gaussian waveguides centered at $x=x_{2i}$ and $x=x_{2i+1}$, respectively. Here, $x_{2i}=-d_{\text{intra}}/2+iL$ and $x_{2i+1}=d_{\text{intra}}/2+iL$, with $i$ being an interger and $L=d_{intra}+d_{inter}$ being the period of the unit cell. In the simulation, by setting $p=6$ and $a=0.5$, we ensure that the waveguides are single-moded. We also set $d_{\text{intra}}=1.8$, $d_{\text{inter}}=2.2$, thus the intra-cell coupling differs from the inter-cell coupling so that a continuous SSH-like model is established.

Equation (\ref{eq10}) is solved using the fourth-order Runge-Kutta algorithm, which is prevalent in solving differential equations \cite{runge}. In our simulation, a step size of $\delta z=0.001$	is employed, and it has been verified that this step size is sufficiently small to ensure the convergence of simulation results. The propagation simulation starts with an excitation condition $\Psi|_{z=0}=\psi_k e^{-\frac{x^8}{\omega^8}}(1+\sigma)$, where $\psi_k$ represents the Bloch mode with Bloch momentum $k$, modulated by a Gaussian envelope of a wide width $\omega=600$. 
Note that we adopted a wide super-Gaussian envelope (containing more than 1000 unit cells) to ensure that the simulation's computation window is sufficiently large such that the light propagation dynamics occuring within the  center of the computationa window is not influenced by the potential boundary effects. $\sigma$ is a random function whose values is arbitrarily selected from the interval $[-1/100,1/100]$, introducing a perturbation strength of $1/100$ into the excitation wavefunction to accelerate the symmetry breaking, if any~\cite{noise}. 

The simulation results of the propagation are given as Fig.~3. Evolution with two different input power per unit cell $P=\int_{\text{unitcell}}|\Psi|^2 dx$ is presented, with one below the predicated $P_{cr}$ (left two columns), and the other at power above $P_{cr}$ (right two columns). Note that, due to variation in the  random noise realization from cell to cell, there are slight power deviations among cells. Therefore, to be specific and without loss of generality, here we focus on the power at the central unit cell. Further, to describe the symmetry breaking, we introduce the partial powers at waveguide $A$ and $B$ from the central cell,  $P_{a}=\int_{-L/2}^{0}|\Psi|^2 dx$ and $P_{b}=\int_{0}^{L/2}|\Psi|^2 dx$. From Fig.~3, it is apparent that the symmetry breaking point significantly drops with the increasing of the Bloch momentum $k$. For instance, at $k=0$, the corresponding Bloch modes, with a power of  $P=0.4$, maintain symmetry over a long distance $z=50$, as shown in  Fig.~3(a) (left panel), as the power distribution between $P_a$ and $P_b$ remains nicely balanced [Fig.~3(a) right panel]. However, once the input power $P$ exceeds $P_{cr}$, such as for the excitation condition at $P=0.6$, power transfer between waveguides occurs after a short distance, as illustrated in Fig.~3(b). Notably, however, in sharp contrast to the monotonic power transfer in a simple double-well potential, where the flow from one site to the other is unidirectional, in cases with multiple double-wells, the power transfer can be reversible. This results in fluctuation in  $P_a$ and $P_b$ with distance $z$. Additionally, due to variations in the initially seeded noise from cell to cell, 
the specific waveguide into which most of the power concentrates varies too.

A similar scenario is observed in the evolution of the Bloch mode at $k=\pi/2L$, but now the symmetry breaking already occurs for an input power of $P=0.37$, as shown in Fig.~3(d). Finally, the minimum power in the symmetry-breaking power is observed for the Bloch wave with $k=\pi/L$: at a power level of $P=0.23$, the symmetry breaking is already evident, as depicted in Fig.~3(f). Therefore, the propagation simulation fully confirmed the aforementioned prediction that, the power threshold for the occurrence of SSB drops with the increasing modulus of quasi-momentum $k$.

\begin{figure*}[htbp]
	\centering
	\includegraphics[width=2\columnwidth]{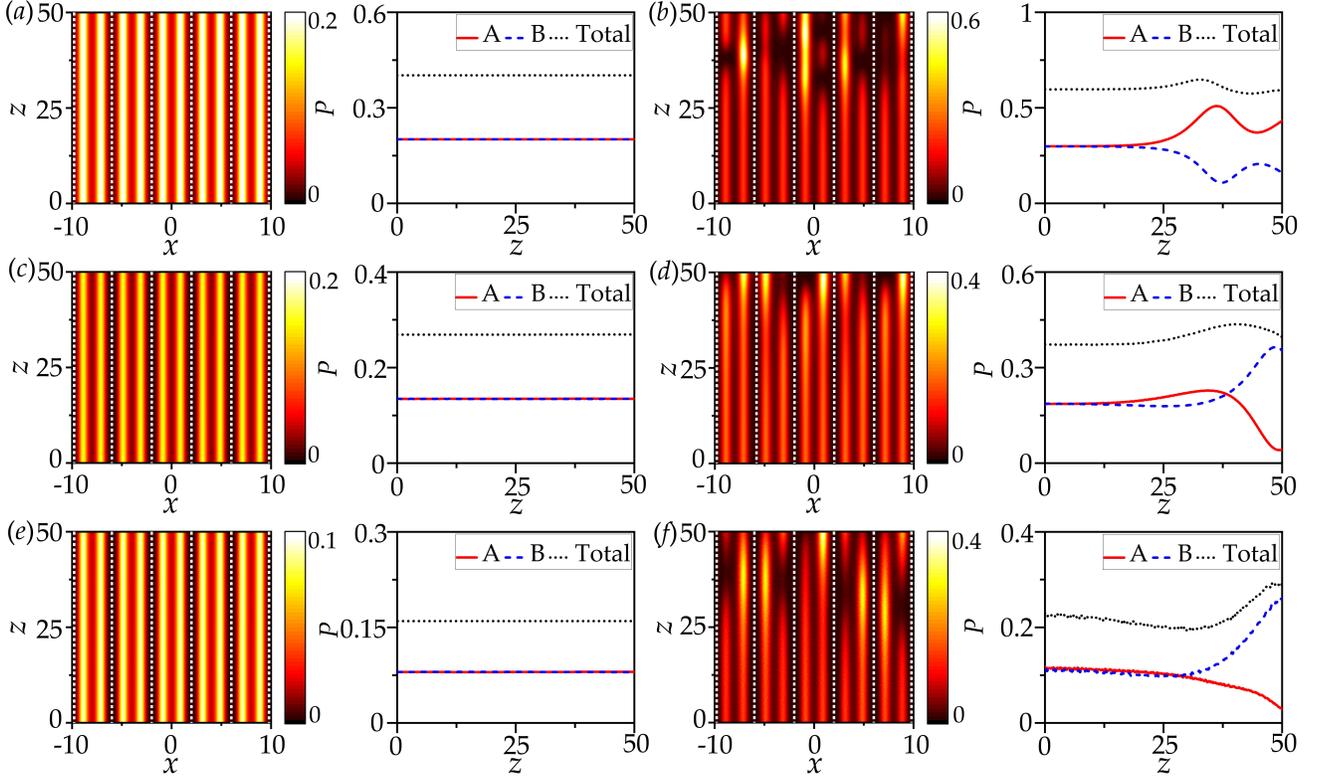}
	\caption {Simulation of light propagation in the 1D SSH system, with input being the symmetric Bloch mode at $k=0$ (first line), $k=\pi/2L$ (second line), and $k=\pi/L$ (third line), respectively. The first and second columns show the evolution of the light field, and the evolution of power at waveguides A and B from the central unit cell when the input power is below their respective $P_{cr}$. The third and fourth columns show the same but with the input power above their respective $P_{cr}$. 
Panels (a) and (b) depict the results for $P=0.4$ and $P=0.6$, respectively, at $k=0$. 
Panels (c) and (d) depict the results for $P=0.27$ and $P=0.37$, respectively, at $k=\pi/2L$. 
Panels (e) and (f) depict the results for $P=0.16$ and $P=0.23$, respectively, at $k=\pi/L$.}
	\label{fig3}
\end{figure*}

\section{\label{sec:level3}Two-dimensional honeycomb lattice model}
\subsection{\label{sec:level3a} Model and Analytical Analysis}

\begin{figure*}[htbp]
	\centering
	\includegraphics[width=2\columnwidth]{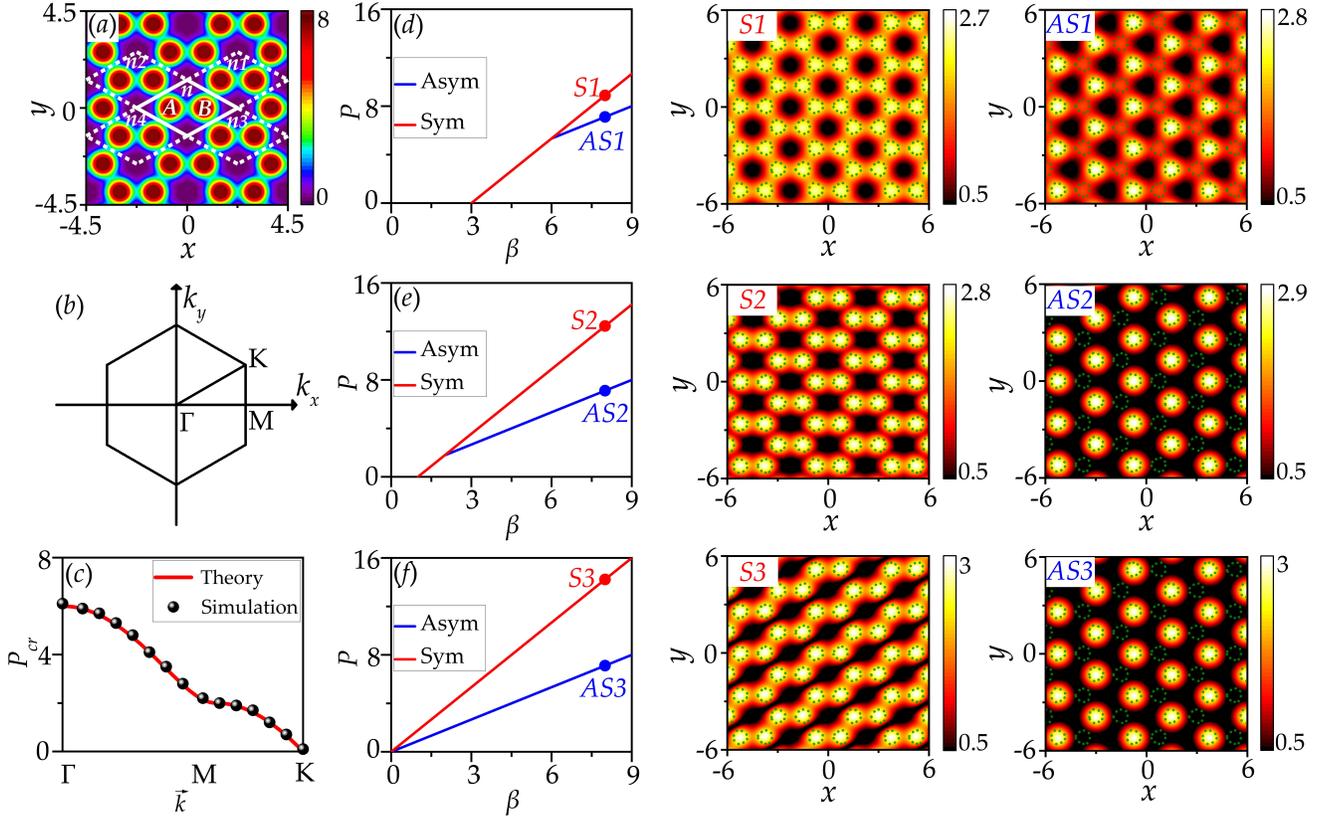}
\caption{Panel (a) illustrates a honeycomb lattice. The white solid line parallelogram represents the $n_{th}$ unit cell of the lattice, while 
the white dashed line parallelogram represents the four neighboured unit cells. In each unit cell, there are two waveguides, identified by $A$ and $B$. (b), the first Brillouin zone, with three high-symmetric points, $\Gamma$, $M$, and $K$ are marked.  (c), the dependence of critical power $P_{cr}$ on quasi-momentum $(k_x, k_y)$. The power curves of symmetric mode (red line) and asymmetric mode (blue line) at $\Gamma$ , $M$ and $K$ point are displayed in panels (d), (e) and (f) respectively.  Examples of the modulus distributions of symmetric (the third column) and asymmetric (the fourth column) mode correspond to the red and blue solid points marked in panels (d), (e) and (f). The green dashed circles indicate the location of the waveguides. }              
	\label{fig4}
\end{figure*}

We here consider the SSB of the nonlinear Bloch modes in a 2D binary periodic system, namely, the honeycomb lattices. As shown in Fig.~\ref{fig4}(a), the honeycomb lattice has two inequivant sites (waveguide $A$ and waveguide $B$) within each unit cell. Within the tightly-binding-approximation (TBA),  the nonlinear coupling equations in honeycomb lattice are given as:
\begin{equation}\label{eq11}
\begin{aligned}
&i\frac{\partial{Q_{a,n}}}{\partial{z}}=-cQ_{b,n}-cQ_{b,n_1}-cQ_{b,n_2}-|Q_{a,n}|^2 Q_{a,n}, \\
&i\frac{\partial{Q_{b,n}}}{\partial{z}}=-cQ_{a,n}-cQ_{a,n_3}-cQ_{a,n_4}-|Q_{b,n}|^2 Q_{b,n}.
\end{aligned}
\end{equation}
Here, $Q_{a,n}$ and $Q_{b,n}$ are the optical field distribution of waveguide $A$ and waveguide $B$ in the $n$th unit cell, respectively.
In this model, there are four nearest unit cells around the $n$th unit cell, named the $n_i (i=1,2,3,4)$ unit cell in proper order.
$Q_{b,n_1}$ and $Q_{b,n_2}$ are the optical field distribution of waveguide $B$ in the $n_1$th and $n_2$th unit cell, while $Q_{a,n_3}$ and $Q_{a,n_4}$ are the optical field distribution of waveguide $A$ in the $n_3$th and $n_4$th unit cell.
$c$ is the coupling factor between waveguides.
According to the Bloch theorem, we consider the Bloch momentum $\bm{k}=(k_x,k_y)$ within the first Brillouin zone [Fig.~\ref{fig4}(b)]. This leads to the expressions $Q_{a,n}=Q_{a,n_i}e^{i\bm{k}\cdot\bm{r_{n,n_i}}} (i=3,4)$ and $Q_{b,n}=Q_{b,n_i}e^{i\bm{k}\cdot\bm{r_{n,n_i}}} (i=1,2)$,  where $\bm{r_{n,n_1}}=(-\frac{3d}{2},-\frac{\sqrt{3}d}{2})$, $\bm{r_{n,n_2}}=(-\frac{3d}{2},\frac{\sqrt{3}d}{2})$, $\bm{r_{n,n_3}}=(\frac{3d}{2},-\frac{\sqrt{3}d}{2})$, and $\bm{r_{n,n_4}}=(\frac{3d}{2},\frac{\sqrt{3}d}{2})$. Here, $d$ represents the  distance between the nearest waveguides $A$ and $B$.

Substituting the above expressions into Eqs.~(\ref{eq11}), we obtain the following equations:
\begin{equation}\label{eq12}
\begin{aligned}
&i\frac{\partial{Q_{a,n}}}{\partial{z}}=g_{1} Q_{b,n}-|Q_{a,n}|^2 Q_{a,n}, \\
&i\frac{\partial{Q_{b,n}}}{\partial{z}}=g_{2} Q_{a,n}-|Q_{b,n}|^2 Q_{b,n}.
\end{aligned}
\end{equation}
Here, $g_{1}=-c-ce^{i\frac{3d}{2} k_x}e^{-i \frac{\sqrt{3}d}{2} k_y}-ce^{i\frac{3d}{2} k_x}e^{i\frac{\sqrt{3}d}{2} k_y}$, and $g_{2}=-c-ce^{-i\frac{3d}{2} k_x}e^{-i\frac{\sqrt{3}d}{2} k_y}-ce^{-i\frac{3d}{2} k_x}e^{i \frac{\sqrt{3}d}{2} k_y}$.
 Stationary nonlinear solutions to Eqs.~(\ref{eq11}) were looked for as $Q_{a,n}=\varphi_{a}e^{i \beta z}$ and $Q_{b,n}=\varphi_{b}e^{i \beta z}$, where the wave functions $\varphi_{a}$ and $\varphi_{b}$ and propagation constant $\beta$ satisfy the following nonlinear equations:
\begin{equation}\label{eq13}
\begin{aligned}
&\beta \varphi_{a}=-g_{1}\varphi_{b}+|\varphi_{a}|^2 \varphi_{a}, \\
&\beta \varphi_{b}=-g_{2}\varphi_{a}+|\varphi_{b}|^2 \varphi_{b}.
\end{aligned}
\end{equation}
Similar to the treatment for 1D systems, see Eq.~(4), in this two-dimensional structure, we assume $\varphi_{a}=\phi_a e^{-i \theta}$ and $\varphi_{b}=\phi_b e^{i \theta}$, respectively, where $\phi_a, \phi_b \in \mathbb{R}$, and $\theta$ is the phase difference of the optical mode between two waveguides. In this way, Eq.~(12) turns into the following equation,  

\begin{equation}\label{eq14}
\begin{aligned}
&\beta \phi_{a}=-g_{1} e^{i2\theta} \phi_{b}+ \phi_{a}^3, \\
&\beta \phi_{b}=-g_{2} e^{-i2\theta} \phi_{a}+ \phi_{b}^3.
\end{aligned}
\end{equation}



We consider the asymmetric mode solution, $\phi_{b}\neq\phi_{a}$, to Eqs.~(\ref{eq14}). 
This solution is readily obtained by setting $-g_{1}e^{i2\theta}=-g_{2}e^{-i2\theta}=C$ (the \textit{effective} coupling coefficient) as Eqs.~(\ref{eq14}) can be recast into the exact 
form of Eqs.~(\ref{eq5}) for 1D binary periodic systems. Thus, apparently, 
the occurrence of the asymmetric mode solution requires $P > 
P_{cr}$, where, cf. Eqs.~(\ref{eq9}),
\begin{equation}\label{eq15}
\begin{aligned}
   P_{cr}&\equiv P_{a}+P_{b} \equiv 2C \\
           &=2c\sqrt{1+4\text{cos}^2\frac{\sqrt3d}{2}k_y+4\text{cos}\frac{3d}{2}k_x\text{cos}\frac{\sqrt3d}{2}k_y}.
\end{aligned}
\end{equation}
Therefore, the critical power $P_{cr}$ that is required to  break the symmetry of the nonlinear Bloch modes is again a function of Bloch momentum $\boldsymbol{k}=(k_x, k_y)$.

The second column of Fig. 4 shows the $P(\beta)$ curve of the 2D symmetric Bloch mode together with the asymmetric modes branching from their symmetric counterparts, at three highest symmetry points, $\Gamma$, i.e., $(k_x, k_y)=(0,0)$ [Fig.~4 (d)], $M$, i.e., $(k_x, k_y)=(\frac{2\pi}{3d},0)$ [Fig.~4 (e)] and $K$, i.e., $(k_x, k_y)=(\frac{2\pi}{3d},\frac{2\pi}{3 \sqrt{3} d})$ [Fig.~4(f)]. Notably, the power at which the branching appears is $P_{cr}=6$ for Bloch modes from the $\Gamma$ point, and $P_{cr}=2$ for Bloch modes from the $M$ point, as well as  $P_{cr}=0$ for Bloch modes from the $K$ point. 
Further, Fig. 4 (c) shows the dependence of $P_{cr}$ on $\boldsymbol{k}$ along the path connecting $\Gamma$, $M$, and $K$. Its monotonic decreasing with the increasing magnitude of wavevector $\boldsymbol{k}$ is obvious. Note that the analytical formula of Eqs.~(\ref{eq15}) is fully collaborated by the result of numerically finding the 2D nonlinear Bloch modes from Eqs.~(\ref{eq13}).  

\subsection{\label{sec:level3c}
Propagation Simulation of SSB in Honeycomb lattices}
Direct propagation simulation in the honeycomb photonic lattices is implemented, based on the following continuous equation,
\begin{equation} \label{eq16}
i \frac{\partial \Psi}{\partial z}=-\frac{1}{2} (\frac{\partial^2 \Psi}{\partial x^2}+\frac{\partial^2 \Psi}{\partial y^2})-V(x,y)\Psi-|\Psi|^2\Psi
\end{equation}
where $V(x,y)=p\sum_{i=1}^{N} e^{-\frac{(x-x_i)^2+(y-y_i)^2}{a^2}}$ describes a 2D arrays of Gaussian waveguides with amplitude $p$ and radius $a$, whose centre $(x_{i},y_{i})$ forms a honeycomb lattice, with $d$ being the distance between the nearest neighbor waveguides. In our simulation, $p=8$, $a=0.5$ and $d=3$.

Similar to the 1D cases, the propagation simulation here is based on the fourth-order Runge-Kutta method. The input beam $\Psi|_{z=0}=\psi_k e^{-\frac{(x^2+y^2)^4}{\omega^8}}(1+\sigma)$. The definition of the Bloch mode $\psi_k$ and random function $\sigma$ are similar to the case in the propagation simulation in 1D binary lattices, however, now they are both 2D functions. Especially, the Bloch wave $\psi_k$ is  characterized by a vector momentum $\boldsymbol{k}$. In the excitation scheme, $\psi_k$ is always chosen from the symmetric branch as shown in Fig.~4(d, e, f), and then we monitor if the input symmetric Bloch function will experience the symmetry broken upon propagation. Also, Similarly to the cases in 1D propagation simulation, the partial power distributed onto waveguide $A$ and $B$ in the central unit cell, $P_a$ and $P_b$, are measured with distance $z$. Here,  
$P_a= \iint_{S_a}|\Psi|^2 dxdy$ and $P_b= \iint_{S_b}|\Psi|^2 dxdy$, where $S_a=\{(x,y)|x<0,y\geq -\frac{\sqrt{3}}{3}x-\frac{\sqrt{3}d}{2},y \leq \frac{\sqrt{3}}{3}x+\frac{\sqrt{3}d}{2}\} $,$S_b=\{(x,y)|x>0,y\geq \frac{\sqrt{3}}{3}x-\frac{\sqrt{3}d}{2},y \leq -\frac{\sqrt{3}}{3}x+\frac{\sqrt{3}d}{2} \} $,
and thus the total power in the central unit cell is $P=P_a+P_b$.

The propagation simulation for Bloch modes from three highest symmetry points $\Gamma, M$ and $K$ are given in Figure 5. For Bloch modes from $\Gamma$ [Fig. 5 (a, b)] and $M$  [Fig. 5 (c, d)], evolution with two slightly different input power $P$ are presented, with one at power that is below the predicated $P_{cr}$, and the other at power above $P_{cr}$.  It is apparent that the symmetry breaking point significantly drops with the increasing of the Bloch momentum $k$. Indeed, for the $\Gamma$ point, the corresponding Bloch modes with a power of $P=0.48$ maintain symmetry over a long distance $z=50$, with power distribution $P_a$ and $P_b$ balanced [Fig.~5(a)]. However, once the input power $P$ is increased above $P_{cr}$, for example, at $P=1.64$, after a short distance, the power at one waveguide starts pumping into the other, quickly resulting in power concentration  at waveguide $B$ [Fig.~5(b)]. A Similar scenario is observed for the evolution of the Bloch mode at the $M$ point, but  symmetry breaking occurs already when the input power is $P=0.33$ [Fig.~5(d)].

Finally, zero-critical power symmetry-breaking is observed for the Bloch wave at the  $K$ point. Even with  extremely low power $P=0.12$, symmetry breaking occurs [Fig.~5(e)]. The symmetry breaking observed in the propagation simulation aligns well with the theoretical and eigenmode analysis. It is also noted that, as noise is randomly distributed from cell to cell, the "gain" or the "lossy" waveguides also varies from waveguide $A$ to $B$ among cells. Additionally, while symmetry broken is typically characterized by an overall decrease or increases in power in a specific waveguide, fluctuations of power in the unit cell are also observed during propagation[Fig.~5(b), lower panel].

Before concluding, we note that the reported symmetry breaking in periodic systems is readily observed in experiments, such as  optical setups  with arrays of waveguides. Let's take for example the 2D setting as described by the dimensionless Eq.~(\ref{eq16}). By referring to specific experimental parameters~\cite{experimentoptica}, the transverse $x$, $y$ and the longitudinal $z$ coordinates are normalized to the characteristic scale $r_0=50~\mathrm{\mu m}$ and diffraction length $kr_0^{2}=14.7~\mathrm{mm}$, respectively. $k=2\pi n/\lambda$ is the wavenumber at light wavelength $\lambda=1550~\mathrm{nm}$, $n\approx 1.45$ is the background refractive index of silica. Under these conditions, a depth of modulation of the refractive index of the waveguide array $p=8$ corresponds to a refractive index contrast of the order of $\delta n \sim 10^{-4}$, dimensionless, and peak intensity $\lvert \Psi \rvert_{max}^2=1$  corresponds to $I \sim 10~\mathrm{GW/cm^2}$  ($n_2 \sim 10^{-20}~\mathrm{m^2/W}$ is the nonlinear refractive index of material)~\cite{experimentoptica}. When the super-Gaussian light is selected as related to the input light as considered in our work, the dimensionless input power $\iint e^{-\frac{(x^2+y^2)^8}{\omega^{16}}}dxdy=1$ corresponds to $P\sim 1~\mathrm{MW}$, which can be readily achieved in the femtosecond-laser experiment~\cite{experimentPRL}.

\begin{figure*}[htbp]
	\centering
	\includegraphics[width=2\columnwidth]{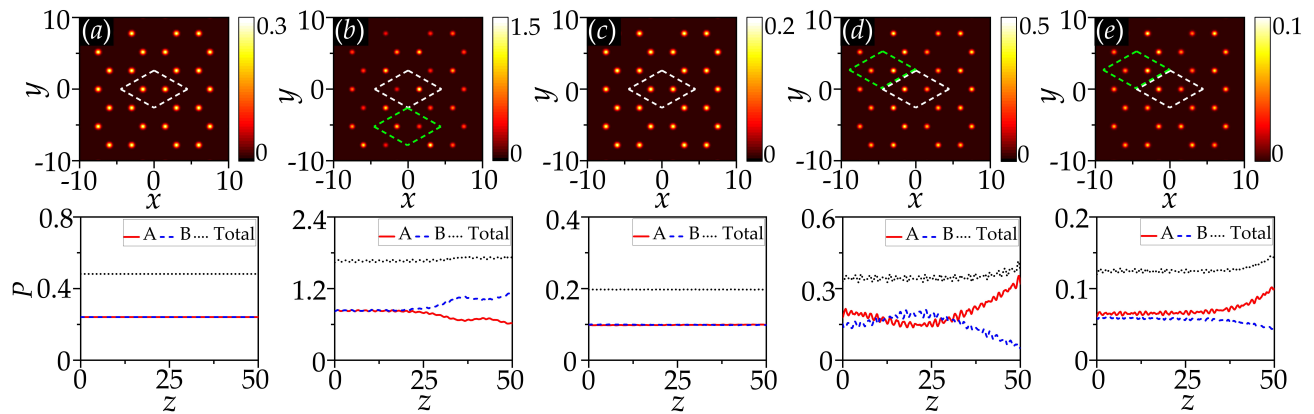}
	\caption{
Simulation of light propagation in the 2D honeycomb system, with input being the symmetric Bloch mode at $(k_x, k_y)=(0,0)$ (a, b),  $(k_x, k_y)=(\frac{2\pi}{3d},0)$ (c, d), and $(k_x, k_y)=(\frac{2\pi}{3d},\frac{2\pi}{3 \sqrt{3} d})$ (e). The first line shows output light amplitude at $z=50$, while the second line shows correspondingly the evolution of the partial power at waveguide $A$ and $B$ from the central unit cell (denoted as a white dashed parallelogram in the first line).
Panels (a) and (b) show the result for  $P=0.48$ and $P=1.64$ at $\Gamma$ point, respectively. 
Panels (c) and (d) show the result for  $P=0.2$ and $P=0.33$ at $M$ point, respectively.
Panels (e) shows the result for  $P=0.12$ at $K$ point. The green dashed parallelogram in panels (b), (d) and (e) shows a different unit cell where the "lossy" and "gain" waveguides are different from the central unit cell.}
	\label{fig5}
\end{figure*}

\section{\label{sec:level4}Conclusion}
Spontaneous symmetry breaking (SSB) is a crucial concept in various nonlinear physics fields, where a symmetric ground state transitions to an asymmetric state under nonlinearity, despite the underlying potential maintaining symmetry. Previous SSB studies focused on one-dimensional finite systems with double- or triple-well potentials, overlooking SSB in periodic systems with binary potentials in unit cells and two-dimensional systems. In our study, we explore SSB in binary periodic systems using the one-dimensional Su-Schrieffer-Heeger (SSH) model and the two-dimensional honeycomb lattice. We derived, within the tightly-binding model, an analytical formula that predicts the power threshold beyond which the symmetry breaking occurs. The analytical analysis is confirmed by the numerical eigen-analysis, as well as the direct propagation simulation using the continuous wave equation. We discovered that the SSB point, where asymmetric nonlinear modes bifurcate from symmetric modes, depends significantly on the Bloch momentum $k$ of the Bloch modes. Specifically, in the SSH model, the critical power for SSB decreases as the Bloch momentum $k$ ranges from $0$ to $\pi/L$. Similarly, in the two-dimensional honeycomb lattice, we observe a similar scenario as $k_x$ and $k_y$ vary from $\Gamma$ to $M$ and then to $K$.  In both systems, Bloch momentum plays a vital role in the SSB of nonlinear Bloch modes in binary periodic lattices, a degree of freedom previously absent in studies of finite optical lattices.




\begin{acknowledgments}
R.P., Q.F. and F.Y. acknowledge support from the NSFC (No.91950120), Scientific funding of Shanghai (No.9ZR1424400), and Shanghai Outstanding Academic Leaders Plan (No.20XD1402000). Q.F. acknowledges 
the support of the National Natural Science Foundation of China (No.12404385) and China Postdoctoral Science Foundation (No.BX20230218, No.2024M751950).
\end{acknowledgments}




\nocite{*}
\thanks{These authors contributed equally to this work.}

\bibliographystyle{unsrt}
\bibliography{apssamp}

\begin{thebibliography}{10}
\bibitem{noise}
In this propagation simulation, a random perturbation is introduced into the incident light beam just once. alternatively, when the random perturbation is continuously applied to the beam throughout its entire propagation period, we observe a similar symmetry-breaking scenario. however, the symmetry-breaking occurs earlier compared to the scenario where the perturbation is introduced only once.
\end{thebibliography}

\end{document}